# Character of the "normal state" of the nickelate superconductors

Kyuho Lee[1,2*], Bai Yang Wang[1,2], Motoki Osada[1,3], Berit H. Goodge[4,5], Tiffany C. Wang[1,6], Yonghun Lee[1,6], Shannon Harvey[1,6], Woo Jin Kim[1,6], Yijun Yu[1,6], Chaitanya Murthy[2], Srinivas Raghu[1,2], Lena F. Kourkoutis[4,5], and Harold Y. Hwang[1,6]

[1] *Stanford Institute for Materials and Energy Sciences, SLAC National Accelerator Laboratory, Menlo Park, CA 94025, USA*
[2] *Department of Physics, Stanford University, Stanford, CA 94305, USA*
[3] *Department of Materials Science and Engineering, Stanford University, Stanford, CA 94305, USA*
[4] *School of Applied and Engineering Physics, Cornell University, Ithaca, NY 14853, USA*
[5] *Kavli Institute at Cornell for Nanoscale Science, Cornell University, Ithaca, NY 14853, USA*
[6] *Department of Applied Physics, Stanford University, Stanford, CA 94305, USA*

## Abstract

The occurrence of superconductivity in proximity to various strongly correlated phases of matter has drawn extensive focus on their normal state properties, to develop an understanding of the state from which superconductivity emerges.[1–4] The recent finding of superconductivity in layered nickelates raises similar interests.[5–9] However, transport measurements of doped infinite-layer nickelate thin films have been hampered by materials limitations of these metastable compounds – in particular, a relatively high density of extended defects.[10–12] Here, by moving to a substrate $(LaAlO_3)_{0.3}(Sr_2TaAlO_6)_{0.7}$ which better stabilizes the growth and reduction conditions, we can synthesize the doping series of $Nd_{1–x}Sr_xNiO_2$ essentially free from extended defects. This enables the first examination of the 'intrinsic' temperature and doping dependent evolution of the transport properties. The normal state resistivity exhibits a low-temperature upturn in the underdoped regime, linear behavior near optimal doping, and quadratic temperature dependence for overdoping. This is strikingly similar to the copper oxides,[2,13] despite key distinctions – namely the absence of an insulating parent compound,[5–7,10,11] multiband electronic structure,[14,15] and a Mott-Hubbard orbital alignment rather than the charge-transfer insulator of the copper oxides.[16,17] These results suggest an underlying universality in the emergent electronic properties of both superconducting families.

---

[*]kyuho@stanford.edu

The idea that superconductivity can arise from doping a correlated insulator has been a pervasive guiding principle since the discovery of the copper oxide superconductors, with impact on materials as far-ranging as twisted bilayer graphene.[1–3] Upon doping the insulator, a 'strange metal' with unconventional electrical transport often occurs and nucleates superconductivity, before further doping gives way to more conventional Fermi-liquid-like behavior. The extent to which this phenomenology requires that the parent compound exhibit a strongly insulating ground state, and whether or not it should exhibit magnetism has been discussed for decades. A further dichotomy – whether the strange metallic behavior reflects the proximity to the correlated insulator or follows from a zero-temperature phase transition to a broken symmetry phase – remains actively debated and largely unresolved.

The observation of superconductivity in a family of layered nickelates presents an opportunity to address some of these perplexing issues. The parent compounds of the infinite-layer nickelates exhibit a weak resistive upturn at low temperatures without a strongly insulating ground state or indications of a gap ($NdNiO_2$ and $PrNiO_2$), and even evidence of superconductivity in $LaNiO_2$.[5–7,10,11] Moreover, to date long-range magnetic order has not been observed in this system.[18–20] Nevertheless, upon doping, we find that $Nd_{1-x}Sr_xNiO_2$ exhibits strange metal behavior with resistivity linearly increasing with temperature $T$ for doping $x$ at the peak of the superconducting dome. Further hole-doping results in a metallic state resembling a Fermi liquid with a resistivity varying as $T^2$, with reduced and ultimately vanishing superconducting scales. Our results suggest therefore that much of the emergent behavior of this class of unconventional superconductors does not strictly require a Mott insulating parent compound with a hard gap to charge excitations. Features in the Hall effect directly correlate with the evolution of the resistivity and may suggest a broken symmetry associated with Fermi surface reconstruction.[2,21,22] Our results can be considered in two contexts – the multiband nature of the electronic structure[14,15,22] and the possibility of a quantum phase transition underlying strange metallic behavior.[2,4]

## Materials advances

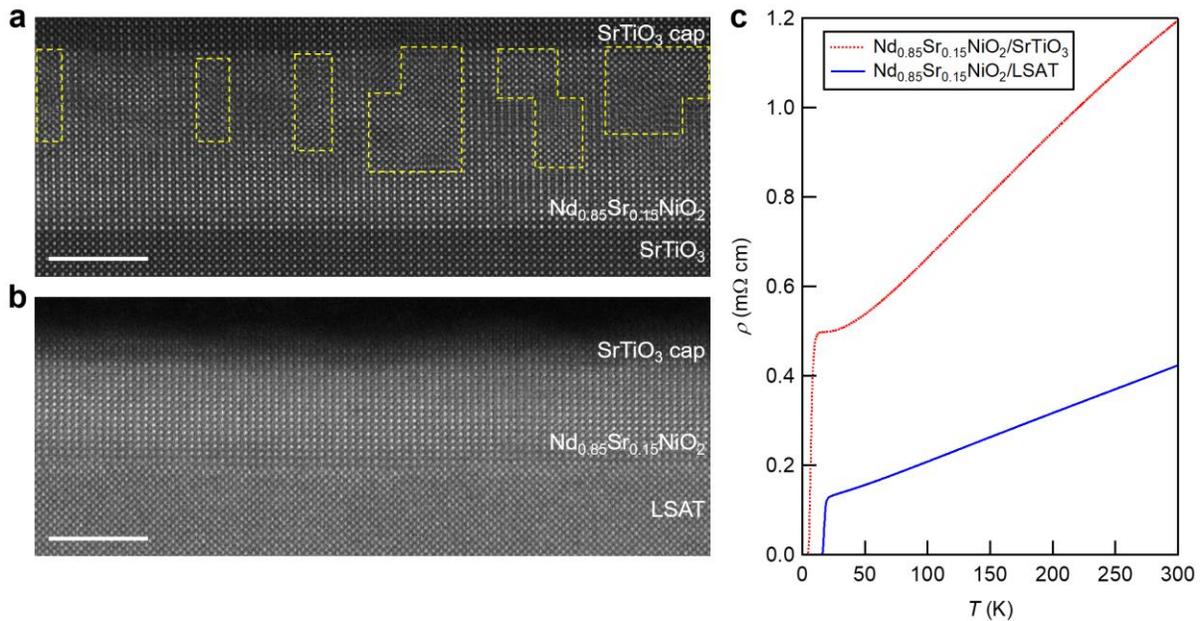

**Fig. 1 | Enhanced crystallinity of Nd$_{1-x}$Sr$_x$NiO$_2$ thin films on LSAT. a,** HAADF-STEM cross-sectional image of a Nd$_{0.85}$Sr$_{0.15}$NiO$_2$ thin film grown on SrTiO$_3$, synthesized using conditions in ref. [10]. Regions bounded by Ruddlesden-Popper type stacking faults are outlined in yellow dashed lines. **b,** HAADF-STEM cross-sectional image of a Nd$_{0.85}$Sr$_{0.15}$NiO$_2$ thin film grown on LSAT, synthesized using the optimized conditions specified in Extended Data Table 1. The film now displays highly uniform crystallinity largely free from extended defects. **c,** Resistivity $\rho$ versus temperature $T$ for the films in **a** (red dotted curve) and **b** (blue solid curve). A significant decrease in the normal-state resistivity reflects the crystallinity improvement. The scale bars are 5 nm.

A central issue for the synthesis and study of superconducting infinite-layer nickelates is materials control due to the poor thermodynamic stability of this system,[10–12] as evident from the orders-of-magnitude variations in the resistivity of infinite-layer nickelates reported across the literature.[10,11,20,23–26] Just as in the development of copper oxides,[1,27,28] minimizing disorder and extrinsic defects is critical for elucidating the intrinsic nature of the normal state and superconducting phase diagram. We have achieved significant advances in the crystallinity of Nd$_{1-x}$Sr$_x$NiO$_2$ ($x$ = 0.05-0.325) by using (LaAlO$_3$)$_{0.3}$(Sr$_2$TaAlO$_6$)$_{0.7}$ (LSAT; lattice constant $a$ = 3.868 Å) substrates to optimize the epitaxial mismatch for both the perovskite precursor and the infinite-layer phases (Extended Data Table 1).[26] As shown in the high-angle annular dark-field (HAADF) scanning transmission electron microscopy (STEM) cross-sectional images (Fig. 1a,b), the Ruddlesden-Popper-type vertical stacking faults[10,12,29] (marked by yellow dashed outlines in Fig. 1a) which densely populate films grown on the widely-used substrate SrTiO$_3$ ($a$ = 3.905 Å) are now essentially eliminated on LSAT, leaving a macroscopically clean thin film with minimal defects (Fig. 1b). This is also reflected in the significant decrease in resistivity $\rho$ (Fig. 1c). Note that the in-plane lattice constant of the films are locked to the substrate in both cases (Fig. 1a,b). X-ray diffraction $\theta$-$2\theta$ symmetric scans show prominent film peaks with out-of-plane lattice constant trends associated with systematic Sr doping (Extended Data Fig. 1). Overall, these data indicate that high crystallinity is uniformly established throughout the probed range of Sr doping, minimizing extended-defect contributions to electrical transport.

## Superconducting dome

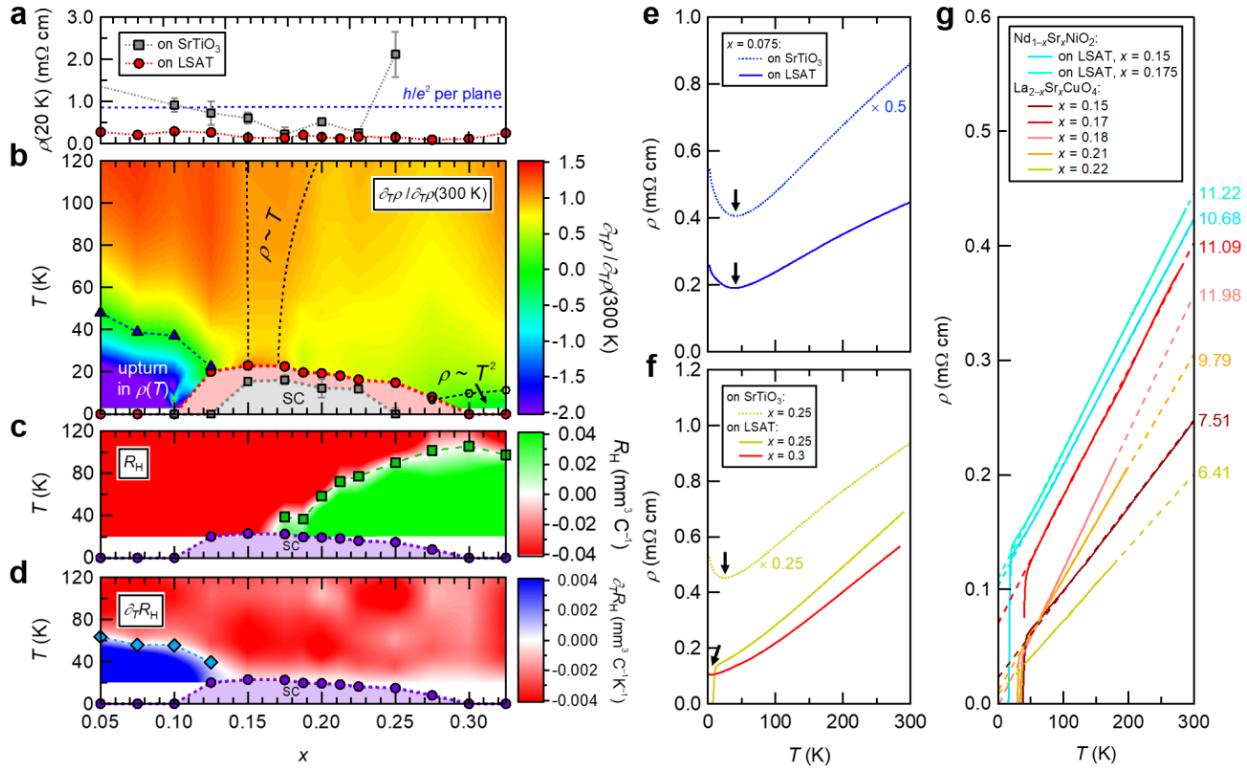

**Fig. 2 | Phase diagram of highly crystalline Nd$_{1-x}$Sr$_x$NiO$_2$. a,** $\rho$(20 K) versus $x$ for Nd$_{1-x}$Sr$_x$NiO$_2$ films on SrTiO$_3$ (ref. [10]) and on LSAT. The LSAT samples uniformly show low $\rho$(20 K) well below the quantum of resistance per NiO$_2$ plane for all $x$. **b,** Contour plot of the slope of $\rho$(T) ($\partial_T\rho$) normalized by the slope at room temperature for the LSAT films. The superconducting dome for films on SrTiO$_3$ (ref. [10]) (grey squares) and here on LSAT (red circles) are also depicted, with the onset transition temperature $T_c$ defined as the temperature at which the second derivative of $\rho$(T) becomes negative. The boundary at which $\partial_T\rho$ is approximately equal to the room-temperature slope (i.e. linear resistivity) is marked by black dotted lines. The temperature-dependence of $\rho$ in the underdoped, optimal, and overdoped regions are labeled. The dark-blue triangles in the underdoped region mark where $\partial_T\rho = 0$ ($T_{upturn}$), while the open circles in the overdoped region mark the temperatures at which the $T^2$ fit deviates (grey arrows in Fig. 4a-c). **c-d,** Contour plots of (**c**) the Hall coefficient $R_H$(T) and (**d**) the slope of $R_H$(T) ($\partial_T R_H$) across the superconducting dome (purple circles). The green squares in **c** show where $R_H$ crosses zero, while the blue diamonds in **d** indicate the temperature at which the local extremum of $R_H$(T) occurs. For **b-d**, the contours were interpolated from the data using natural neighbor interpolation. **e-g,** $\rho$(T) at representative $x$ characteristic of the (**e**) underdoped, (**f**) overdoped, and (**g**) optimally doped regions. Dotted curves are for films grown on SrTiO$_3$,[10] while solid curves are the optimized films grown on LSAT. Black arrows in **e** and **f** indicate $T_{upturn}$ at which the resistive upturn occurs. For **g**, $\rho$(T) of single-crystal La$_{2-x}$Sr$_x$CuO$_4$ with $T$-linear normal-state $\rho$ are also plotted for comparison.[30–32] The dashed lines are linear fits to the normal-state $\rho$, with the slopes of the linear fit indicated at the right in units of $10^{-4}$ mΩ cm K$^{-1}$.

This optimized sample series on LSAT provides a robust platform to investigate the inherent phase diagram of the infinite-layer nickelates. We first observe that $\rho$(20 K) now maintains a similar range of 0.1-0.3 m$\Omega$ cm across all $x$, significantly below the scale of a resistance quantum $R_q$ per NiO$_2$ plane (Fig. 2a; see Extended Data Fig. 2 for all individual $\rho(T)$ curves). Importantly, this includes the underdoped and overdoped regimes, with $\rho$(20 K) approximately 5~30 times lower than any previously reported (~0.6-9 m$\Omega$ cm).[5,10–12,23] This suggests that the low-temperature normal-state scattering rate is comparable across the phase diagram, although multiband effects should be considered.

It is noteworthy that a superconducting dome is still observed for films on LSAT (Fig. 2b). For films on SrTiO$_3$, a direct correlation was found between the presence of superconductivity and the magnitude of the normal-state resistance with respect to $R_q$.[7,10] The fact that the normal state resistance depends significantly on the substrate (LSAT vs. SrTiO$_3$), while the existence of the superconducting dome does not, confirms that the dome itself is an intrinsic property of these doped nickelates. However, at a quantitative level, the superconducting dome in the lower resistance films on LSAT is significantly larger, with onset transition temperature $T_c$ above 20 K for optimal doping at $x \approx 0.15$-$0.175$ (Fig. 2b), and with a width $\Delta x \approx 0.2$ that is now very comparable to that of the copper oxides ($\Delta x \approx 0.21$ for La$_{2-x}$Sr$_x$CuO$_4$)[13] (Extended Data Fig. 3). Both the robustness of the dome and the higher $T_c$ indicate that superconductivity here must be unconventional and cannot be explained purely by electron-phonon mechanisms.[33]

**Normal-state phase diagram**

Figure 2b shows the variation of the slope of $\rho(T)$ normalized to the room temperature value across the phase diagram. Three regimes of behavior are observed, with representative data shown in Fig. 2e-g: a resistive upturn in the underdoped region characterized by $T_{\text{upturn}}$, the temperature at which the resistivity minimum occurs ($d\rho/dT = 0$); $\rho \sim T^2$ in the overdoped region; and a narrow range of $\rho \sim T$ at the peak of the superconducting dome.

The evolution of the resistivity is accompanied by systematic features in the Hall coefficient $R_H$. At high temperatures and low doping $R_H$ is negative, while it is positive in the low temperature limit beyond optimal doping (Fig. 2c; see Extended Data Fig. 4 for all individual $R_H(T)$ curves).

The boundary defining the sign change in $R_H$ extrapolates to $T = 0$ at optimal doping. The second clear feature in $R_H$ can be seen in the underdoped region. Here, $R_H$ is negative at all temperatures and shows a pronounced local maximum (Fig. 2d). The temperature at which this maximum occurs decreases as a function of doping and tracks the resistive upturn (dark blue triangles in Fig. 2b), such that both features extrapolate to vanish under the peak of the superconducting dome.

**Underdoped regime**

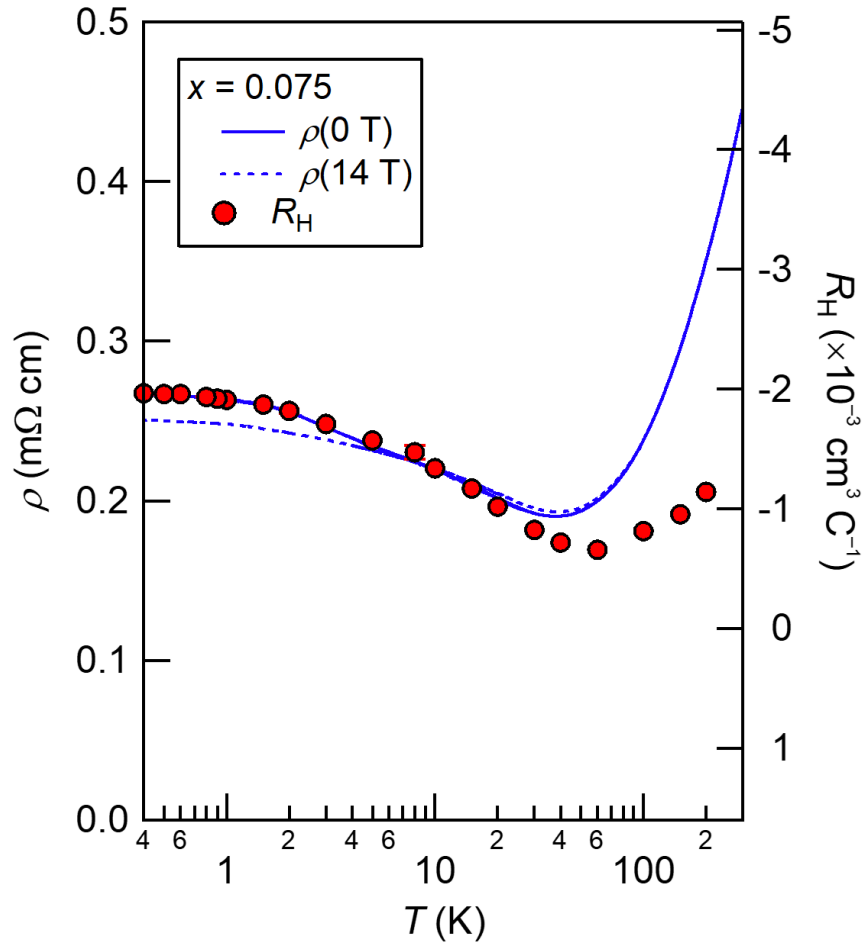

**Fig. 3 | Magnetotransport characteristics of the underdoped regime.** Temperature dependence of $\rho$ (left; at 0 and 14 T) and $R_H$ (right; note the reverse vertical scale) for $x = 0.075$. The functional form of $\rho$ and $R_H$ converges as the resistive upturn onsets at lower temperatures. At $T < 5$ K, $\rho$ deviates from a $\sim \log(1/T)$ upturn and saturates instead, with the saturation also observed in $R_H$.

In this region, the low-temperature resistivity has been noted to vary as $\sim \log(1/T)$ for films on $SrTiO_3$.[34] Comparing data for films on LSAT, we see that although the resistivity itself is a factor of ~4 smaller in the cleaner samples, $T_{upturn}$ is nearly the same (Fig. 2e). $T_{upturn}$ is also essentially constant as a function of magnetic field, with an overall relatively small magnetoresistance (Fig. 3). These observations suggest that the resistive upturn cannot be directly ascribed to disorder or localization effects (with or without interaction corrections), nor to Kondo physics.[35] Indeed, by extending measurements down to lower temperatures, we observe a saturation in resistivity below 2 K (Fig. 3).

Both the resistive upturn and ultimate saturation are closely tracked by $R_H(T)$, with identical functional form (Fig. 3). This behavior is naturally and most simply explained within a two-band model with both electron and hole carriers. Electronic structure calculations for the nickelates indicate the presence of a large hole pocket with $3d_{x^2-y^2}$ character, and electron pockets arising from Nd 5$d$ and Ni 3$d$ hybridization, making this an intrinsically multiband system.[14,15,22] A two-band analysis concludes that the observed magnetotransport behavior corresponds to a decrease in the hole carrier contribution in the presence of parallel electron conduction (see Supplementary Information).

The relative insensitivity of $T_{upturn}$ to disorder strength (Fig. 2e) further suggests that the upturn is caused by intrinsic strong correlation effects that effectively freeze out the hole contribution to conduction. The very recent observations of charge order in infinite-layer nickelates with an incommensurate wave vector of (~1/3, 0) reciprocal lattice units[36–38] provide a candidate for the correlations driving the upturn. This would also be consistent with the lack of magnetic field dependence of $T_{upturn}$. Notably, the upturn in resistivity – with $T_{upturn}$ decreasing as a function of $x$ – and the presence of strong correlation effects, is highly reminiscent of the underdoped region of the copper oxide phase diagram (Extended Data Fig. 3).[2,13,30,31]

## Overdoped regime

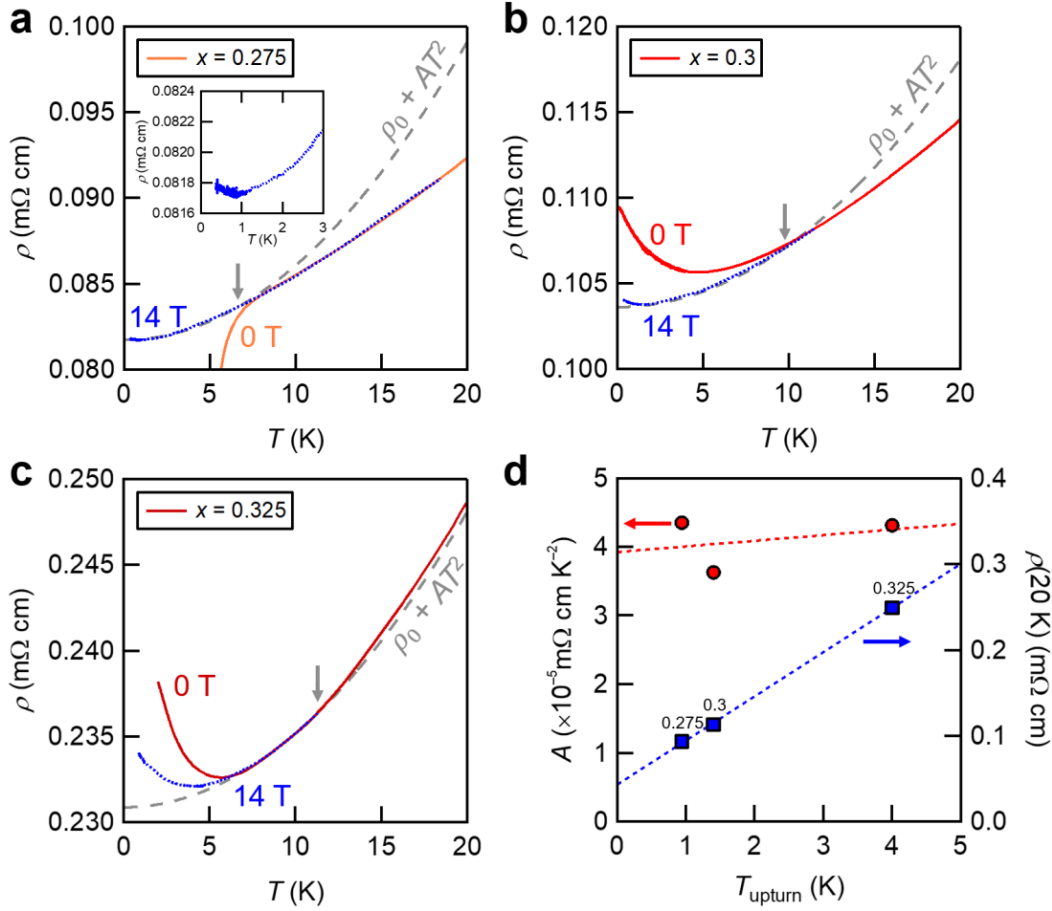

**Fig. 4 | Magnetotransport characteristics of the overdoped regime. a-c,** $\rho$ versus $T$ for overdoped $Nd_{1-x}Sr_xNiO_2$ films, with (**a**) $x = 0.275$, (**b**) $x = 0.3$, and (**c**) $x = 0.325$. Solid curves are measured at zero magnetic field, while dotted blue curves are measured in 14 T applied magnetic field. The $\rho_0 + AT^2$ quadratic fits are shown as dashed grey curves. The grey arrows indicate the temperature at which the quadratic fit deviates. The inset in **a** shows a magnified view of the small resistive upturn at $T < 1$ K under 14 T applied magnetic field. **d,** Quadratic fit parameter $A$ (left, red circles) and $\rho(20\ K)$ (right, blue squares) versus $T_{upturn}$ at 14 T for the three films in **a-c**, with linear fits shown as dashed lines. $A$ shows minimal correlation with $T_{upturn}$, while $\rho(20\ K)$ shows a strong linear trend with $T_{upturn}$.

By contrast to the underdoped regime, the temperature onset of the resistive upturn in the overdoped region is considerably reduced upon lowering disorder (Fig. 2f) and under applied magnetic field (Fig. 4a-c). Indeed, $T_{upturn}$ shows a strong linear correlation with the low-temperature normal-state resistivity (Fig. 4d). These observations indicate that the resistive upturn in the overdoped region is driven by disorder physics, unlike in the underdoped region where, as we have discussed, it appears to arise from correlation effects. Suppressing the upturn (or superconductivity) with a 14 T magnetic field reveals metallic Fermi-liquid-like $\rho \sim AT^2$ at low temperatures, with the quadratic coefficient $A$ essentially uncorrelated with $T_{upturn}$ or $x$ (Fig. 4d).

In addition, the temperature at which the measured resistivity starts to deviate from the low-temperature $T^2$ fit increases as a function of $x$ (Fig. 4a-c), in a manner analogous to the crossover in the functional form of $\rho$ from $T + T^2$ or $T^n$ ($1 < n < 2$) to $T^2$ in the overdoped region of the copper oxide phase diagram.[2,39,40] Furthermore, $R_H(T)$ rather resembles that in overdoped $La_{2-x}Sr_xCuO_4$ in terms of magnitude, functional form, and a sign change at ~100 K.[41,42]

## *T*-linear resistivity at optimal doping

A notable feature that emerges for the highly crystalline films on LSAT is the strange metallic $\rho \sim T$ behavior in the normal state near optimal doping of $x \approx 0.15$-$0.175$ (Fig. 2b,g). The evolution of the functional form and the emergence of *T*-linear behavior upon reduced disorder (Fig. 1c) are similarly observed in disorder studies of the copper oxides.[43,44] Interestingly, the *T*-linear slope of the infinite-layer nickelates (~11 mΩ cm K$^{-1}$) is remarkably close to that of $La_{2-x}Sr_xCuO_4$ (~6-11 mΩ cm K$^{-1}$).[30–32] Among the copper oxides, this comparison appears most apt given that both systems share the solid-solution cation disorder associated with chemical hole doping. The linear resistivity in copper oxides has been a longstanding puzzle, and it is one of several examples that have raised consideration of fundamental bounds on scattering rates.[45,46] For the nickelates, the lack of experimental measures of the multiband effective masses and carrier densities preclude attempts at a quantitative analysis. However, it is surprising that the value of the nickelate *T*-linear slope itself is so close to that in $La_{2-x}Sr_xCuO_4$ despite the significant differences in the electronic structure which result in parallel conduction channels. While the ultimate origins of strange metallicity (and the upturn in resistance on the underdoped side) are yet unclear, the observations here directly indicate that a parent compound with a hard insulating gap is not a crucial ingredient for the strange metal physics that ensues near optimal doping for superconductivity.

## Discussion

If strange metallicity is not directly tied to a proximate Mott insulator, a commonly invoked alternative would involve scattering off of soft order parameter fluctuations associated with a continuous quantum phase transition occluded by the superconducting dome.[2,4,47] Further studies are needed to determine the nature of such fluctuations if present – the charge stripe phase is a plausible candidate[36–38] – and their effect on the resistive upturn in the underdoped regime. When all of the magnetotransport features are summarized on a common phase diagram (Extended Data Fig. 5), it is suggestive of such a quantum critical scenario: the monotonic decrease of the resistive $T_{upturn}$ (and associated $R_H$ maximum) with doping and its apparent vanishing near optimal doping; and Hall measurements which mark a locus of doping dependent temperatures on the overdoped side where the effective sign of the carriers changes, which again vanishes approaching optimal doping. These features have been presented in a two-band picture appropriate for the nickelates, and strictly within this framework the evolution of $R_H$ could be coincidental, reflecting the details of the underlying fermiology of the material – there are analogous debates on multiband effects in the copper oxides, for both hole and electron doping.[21] Alternatively, the vanishing Hall number could be ascribed to Fermi surface reconstruction associated with a density wave order parameter for both nickelates and copper oxides, and thus of more fundamental significance.

With a resistive upturn in the underdoped region driven by strong electron correlations, a non-Fermi-liquid $T$-linear resistivity near optimal doping, and a Fermi-liquid ground state in the overdoped region, the superconducting phase diagram of the infinite-layer nickelates bears a remarkable resemblance to that of the copper oxides (Extended Data Fig. 5). This is quite surprising, especially considering the key distinctions between the two systems. The undoped parent state of the nickelates is not an antiferromagnetic insulator.[7,10,11,18–20] The hybridization between the Nd 5$d$ and Ni 3$d$ bands introduces additional electron pockets in the Fermi surface, making the nickelates an intrinsically multiband system.[14,15,22] Spectroscopic measurements suggest that the orbital alignment of the nickelates is closer to the Mott-Hubbard regime, rather than the charge-transfer regime like the hole-doped copper oxides.[16,17] And yet, the nickelate phase diagram – in particular the superconducting dome and the electrical transport in the normal state – is so similar to that of the copper oxides. This phenomenology extends beyond these oxides to materials as disparate as twisted bilayer graphene and chalcogenides,[3,4,48] indicating an underlying universality in their electrical transport that remains to be understood.

## Methods

**Film growth:** Polycrystalline $Nd_{1-x}Sr_xNi_{1.15}O_3$ targets ($x$ = 0.05-0.325) were prepared by pelletizing mixtures of $Nd_2O_3$, $SrCO_3$, and NiO powders, decarbonating at 1200 °C for 12 hours, regrinding and re-pelletizing, and then sintering at 1350 °C for 12 hours.[5,12] ~15 unit cells of $Nd_{1-x}Sr_xNiO_3$ epitaxial thin films were grown by pulsed-laser deposition with a KrF excimer laser ($\lambda$ = 248 nm) on 5 × 5 mm$^2$ LSAT (001) substrates, with the substrate surface prepared *ex situ* by standard acetone-isopropyl alcohol ultrasonication. The films were synthesized using the conditions specified in Extended Data Table 1. ~4 unit cells of $SrTiO_3$ (001) were grown *in situ* as a capping layer.[12]

**Reduction process:** After cutting into two 2.5 × 5 mm$^2$ pieces, the perovskite samples were vacuum-sealed (< 0.1 mTorr) with ~0.1 g of $CaH_2$ powder in a Pyrex glass tube, loosely wrapped with aluminum foil to avoid direct contact with $CaH_2$. The glass tube was first heated at 260 °C for 2 hours, with temperature ramp rate of 10 °C min$^{-1}$. Then, x-ray diffraction (XRD) $\theta$-$2\theta$ symmetric scan and $\rho(T)$ measurements were performed *ex situ* to evaluate the degree of topotactic transition. 30-minute reductions at 260 °C and *ex situ* characterizations were incrementally continued until the out-of-plane lattice constant, superconducting transition temperature, and residual resistivity ratio saturated, indicating complete reduction. The total reduction time across all samples was ~2.5-4 hours.

**Characterization:** XRD $\theta$-$2\theta$ symmetric scans were measured using a monochromated Cu $K_{\alpha 1}$ source ($\lambda$ = 1.5406 Å). Cross-sectional STEM specimens were prepared by a standard focused ion beam (FIB) lift-out process on a Thermo Scientific Helios G4 X FIB. HAADF-STEM images of the specimens were acquired on an aberration-corrected Thermo Fisher Scientific Spectra 300 X-CFEG operated at 300 kV with a probe convergence semi-angle of 30 mrad and inner (outer) collection angles of 66 (200) mrad. The measurements of $\rho(H, T)$ and Hall effect were conducted in a six-point Hall bar geometry using aluminum wire-bonded contacts. The Hall effect was measured to be linear up to the highest measured magnetic field of 14 T.

## Data availability

The data that support the findings of this study are available from the corresponding author upon request.

## Acknowledgements

We thank Steve Kivelson, Tom Devereaux, and Marc Gabay for useful discussions. This work was supported by the U. S. Department of Energy, Office of Basic Energy Sciences, Division of Materials Sciences and Engineering (Contract No. DE-AC02-76SF00515) and the Gordon and Betty Moore Foundation's Emergent Phenomena in Quantum Systems Initiative (Grant No. GBMF9072, synthesis equipment). C.M. acknowledges support by the Gordon and Betty Moore Foundation's Emergent Phenomena in Quantum Systems Initiative (Grant No. GBMF8686). B.H.G. and L.F.K. acknowledge support by the Department of Defense Air Force Office of Scientific Research (Grant No. FA 9550-16-1-0305) and the Packard Foundation. This work made use of a Helios FIB supported by the NSF (Grant No. DMR-1539918) and the Cornell Center for Materials Research Shared Facilities, which are supported through the NSF MRSEC program (Grant No. DMR-1719875). The Thermo Fisher Spectra 300 X-CFEG was acquired with support from Platform for the Accelerated Realization, Analysis, and Discovery of Interface Materials (PARADIM), an NSF MIP (No. DMR-2039380) and Cornell University.

## Author contributions

K.L. and H.Y.H. conceived the project. K.L. and Y.L. fabricated the polycrystalline targets. K.L. fabricated the perovskite thin films. M.O., Y.L., and W.J.K. performed the soft-chemistry reductions. K.L. conducted x-ray diffraction characterizations. B.Y.W., T.C.W., Y.L., S.H., W.J.K., and Y.Y. performed the transport measurements. B.H.G. and L.F.K. conducted STEM measurements. K.L., B.Y.W., C.M., S.R., and H.Y.H. wrote the manuscript with input from all authors.

## Competing interests

The authors declare no competing interests.

# References

1. Lee, P. A., Nagaosa, N. & Wen, X. G. Doping a Mott Insulator: Physics of High-Temperature Superconductivity. *Rev. Mod. Phys.* **78**, 17 (2006).
2. Keimer, B., Kivelson, S. A., Norman, M. R., Uchida, S. & Zaanen, J. From Quantum Matter to High-Temperature Superconductivity in Copper Oxides. *Nature* **518**, 179 (2015).
3. Cao, Y. *et al.* Unconventional Superconductivity in Magic-Angle Graphene Superlattices. *Nature* **556**, 43 (2018).
4. Fernandes, R. M. *et al.* Iron Pnictides and Chalcogenides: a New Paradigm for Superconductivity. *Nature* **601**, 35 (2022).
5. Li, D. *et al.* Superconductivity in an Infinite-Layer Nickelate. *Nature* **572**, 624 (2019).
6. Osada, M. *et al.* A Superconducting Praseodymium Nickelate with Infinite Layer Structure. *Nano Lett.* **20**, 5735 (2020).
7. Osada, M. *et al.* Nickelate Superconductivity without Rare-Earth Magnetism: (La,Sr)NiO$_2$. *Adv. Mater.* **2021**, 2104083 (2021).
8. Pan, G. A. *et al.* Superconductivity in a Quintuple-Layer Square-Planar Nickelate. *Nat. Mater.* **21**, 160 (2022).
9. Zeng, S. *et al.* Superconductivity in Infinite-Layer Nickelate La$_{1-x}$Ca$_x$NiO$_2$ Thin Films. *Sci. Adv.* **8**, eabl9927 (2022).
10. Li, D. *et al.* Superconducting Dome in Nd$_{1-x}$Sr$_x$NiO$_2$ Infinite Layer Films. *Phys. Rev. Lett.* **125**, 027001 (2020).
11. Zeng, S. *et al.* Phase Diagram and Superconducting Dome of Infinite-Layer Nd$_{1-x}$Sr$_x$NiO$_2$ Thin Films. *Phys. Rev. Lett.* **125**, 147003 (2020).
12. Lee, K. *et al.* Aspects of the Synthesis of Thin Film Superconducting Infinite-Layer Nickelates. *APL Mater.* **8**, 041107 (2020).
13. Takagi, H. *et al.* Superconductor-to-Nonsuperconductor Transition in (La$_{1-x}$Sr$_x$)$_2$CuO$_4$ as Investigated by Transport and Magnetic Measurements. *Phys. Rev. B* **40**, 2254 (1989).
14. Lee, K. W. & Pickett, W. E. Infinite-Layer LaNiO$_2$: Ni$^{1+}$ is not Cu$^{2+}$. *Phys. Rev. B* **70**, 165109 (2004).
15. Botana, A. S. & Norman, M. R. Similarities and Differences between LaNiO$_2$ and CaCuO$_2$ and Implications for Superconductivity. *Phys. Rev. X* **10**, 011024 (2020).
16. Hepting, M. *et al.* Electronic Structure of the Parent Compound of Superconducting Infinite-Layer Nickelates. *Nat. Mater.* **19**, 381 (2020).
17. Goodge, B. H. *et al.* Doping Evolution of the Mott-Hubbard Landscape in Infinite-Layer Nickelates. *Proc. Natl. Acad. Sci. U. S. A.* **118**, e2007683118 (2021).
18. Hayward, M. A., Green, M. A., Rosseinsky, M. J. & Sloan, J. Sodium Hydride as a Powerful Reducing Agent for Topotactic Oxide Deintercalation: Synthesis and Characterization of the Nickel(I) Oxide LaNiO$_2$. *J. Am. Chem. Soc.* **121**, 8843 (1999).
19. Hayward, M. A. & Rosseinsky, M. J. Synthesis of the Infinite Layer Ni(I) Phase NdNiO$_{2+x}$ by Low Temperature Reduction of NdNiO$_3$ with Sodium Hydride. *Solid State Sci.* **5**, 839 (2003).
20. Wang, B.-X. *et al.* Synthesis and Characterization of Bulk Nd$_{1-x}$Sr$_x$NiO$_2$ and Nd$_{1-x}$Sr$_x$NiO$_3$. *Phys. Rev. Mater.* **4**, 084409 (2020).
21. Greene, R. L., Mandal, P. R., Poniatowski, N. R. & Sarkar, T. The Strange Metal State of the Electron-Doped Cuprates. *Annu. Rev. Condens. Matter Phys.* **11**, 213 (2020).
22. Leonov, I., Skornyakov, S. L. & Savrasov, S. Y. Lifshitz Transition and Frustration of Magnetic Moments in Infinite-Layer NdNiO$_2$ upon Hole Doping. *Phys. Rev. B* **101**, 241108(R) (2020).

23. Gu, Q. *et al.* Single Particle Tunneling Spectrum of Superconducting $Nd_{1-x}Sr_xNiO_2$ Thin Films. *Nat. Commun.* **11**, 6027 (2020).
24. Li, Y. *et al.* Impact of Cation Stoichiometry on the Crystalline Structure and Superconductivity in Nickelates. *Front. Phys.* **9**, 719534 (2021).
25. Gao, Q., Zhao, Y., Zhou, X. J. & Zhu, Z. Preparation of Superconducting Thin Films of Infinite-Layer Nickelate $Nd_{0.8}Sr_{0.2}NiO_2$. *Chinese Phys. Lett.* **38**, 077401 (2021).
26. Ren, X. *et al.* Superconductivity in Infinite-Layer $Pr_{0.8}Sr_{0.2}NiO_2$ Films on Different Substrates. *arXiv:2109.05761* (2021).
27. Attfield, J. P., Kharlanov, A. L. & McAllister, J. A. Cation Effects in Doped $La_2CuO_4$ Superconductors. *Nature* **394**, 157 (1998).
28. Kim, G. *et al.* Optical Conductivity and Superconductivity in Highly Overdoped $La_{2-x}Ca_xCuO_4$ Thin Films. *Proc. Natl. Acad. Sci. U. S. A.* **118**, e2106170118 (2021).
29. Guo, Q., Farokhipoor, S., Magén, C., Rivadulla, F. & Noheda, B. Tunable Resistivity Exponents in the Metallic Phase of Epitaxial Nickelates. *Nat. Commun.* **11**, 2949 (2020).
30. Takagi, H. *et al.* Systematic Evolution of Temperature-Dependent Resistivity in $La_{2-x}Sr_xCuO_4$. *Phys. Rev. Lett.* **69**, 2975 (1992).
31. Boebinger, G. S. *et al.* Insulator-to-Metal Crossover in the Normal State of $La_{2-x}Sr_xCuO_4$ Near Optimum Doping. *Phys. Rev. Lett.* **77**, 5417 (1996).
32. Cooper, R. A. *et al.* Anomalous Criticality in the Electrical Resistivity of $La_{2-x}Sr_xCuO_4$. *Science* **323**, 603 (2009).
33. Nomura, Y. *et al.* Formation of a Two-Dimensional Single-Component Correlated Electron System and Band Engineering in the Nickelate Superconductor $NdNiO_2$. *Phys. Rev. B* **100**, 205138 (2019).
34. Zhang, G. M., Yang, Y. F. & Zhang, F. C. Self-Doped Mott Insulator for Parent Compounds of Nickelate Superconductors. *Phys. Rev. B* **101**, 020501(R) (2020).
35. Hsu, Y. T. *et al.* Insulator-to-Metal Crossover Near the Edge of the Superconducting Dome in $Nd_{1-x}Sr_xNiO_2$. *Phys. Rev. Res.* **3**, L042015 (2021).
36. Rossi, M. *et al.* A Broken Translational Symmetry State in an Infinite-Layer Nickelate. *arXiv:2112.02484* (2021).
37. Krieger, G. *et al.* Charge and Spin Order Dichotomy in $NdNiO_2$ Driven by $SrTiO_3$ Capping Layer. *arXiv:2112.03341* (2021).
38. Tam, C. C. *et al.* Charge Density Waves in Infinite-Layer $NdNiO_2$ Nickelates. *arXiv:2112.04440* (2021).
39. Hussey, N. E. Phenomenology of the Normal State In-Plane Transport Properties of High-$T_c$ Cuprates. *J. Phys.: Condens. Matter* **20**, 123201 (2008).
40. Jin, K., Butch, N. P., Kirshenbaum, K., Paglione, J. & Greene, R. L. Link Between Spin Fluctuations and Electron Pairing in Copper Oxide Superconductors. *Nature* **476**, 73 (2011).
41. Hwang, H. Y. *et al.* Scaling of the Temperature Dependent Hall Effect in $La_{2-x}Sr_xCuO_4$. *Phys. Rev. Lett.* **72**, 2636 (1994).
42. Ando, Y., Kurita, Y., Komiya, S., Ono, S. & Segawa, K. Evolution of the Hall Coefficient and the Peculiar Electronic Structure of the Cuprate Superconductors. *Phys. Rev. Lett.* **92**, 197001 (2004).
43. Fukuzumi, Y., Mizuhashi, K., Takenaka, K. & Uchida, S. Universal Superconductor-Insulator Transition and $T_c$ Depression in Zn-Substituted High-$T_c$ Cuprates in the Underdoped Regime. *Phys. Rev. Lett.* **76**, 684 (1996).

44. Rullier-Albenque, F., Alloul, H. & Tourbot, R. Disorder and Transport in Cuprates: Weak Localization and Magnetic Contributions. *Phys. Rev. Lett.* **87**, 157001 (2001).
45. Gurvitch, M. & Fiory, A. T. Resistivity of $La_{1.825}Sr_{0.175}CuO_4$ and $YBa_2Cu_3O_7$ to 1100 K: Absence of Saturation and Its Implications. *Phys. Rev. Lett.* **59**, 1337 (1987).
46. Bruin, J. A. N., Sakai, H., Perry, R. S. & Mackenzie, A. P. Similarity of Scattering Rates in Metals Showing *T*-Linear Resistivity. *Science* **339**, 804 (2013).
47. Taillefer, L. Scattering and Pairing in Cuprate Superconductors. *Annu. Rev. Condens. Matter Phys.* **1**, 51 (2010).
48. Ghiotto, A. *et al.* Quantum Criticality in Twisted Transition Metal Dichalcogenides. *Nature* **597**, 345 (2021).

**Extended Data Table 1.** Pulsed-laser deposition growth parameters for perovskite $Nd_{1-x}Sr_xNiO_3$ thin films on LSAT. The optimized infinite-layer films were obtained across $x$ by reducing the perovskite films at 260 °C for ~2.5-4 hours (see Methods). We note that under the same reduction conditions, complete reduction could not be achieved for $x < 0.05$, and thus we limit our study to $x$ = 0.05-0.325.

| Growth Parameters | Parameter Values |
|---|---|
| Fluence (J cm$^{-2}$) | 2.6 |
| Laser spot size (mm$^2$) | 0.77 |
| Target composition | $Nd_{1-x}Sr_xNi_{1.15}O_3$ |
| Pre-ablation pulse number | 241 |
| Target track outer diameter (cm) | 1.0-1.1 |
| Substrate-to-Target Distance (mm) | 55.5 |
| $P_{O_2}$, deposition (mTorr) | 150 |
| $T$, deposition (°C) | 580 |

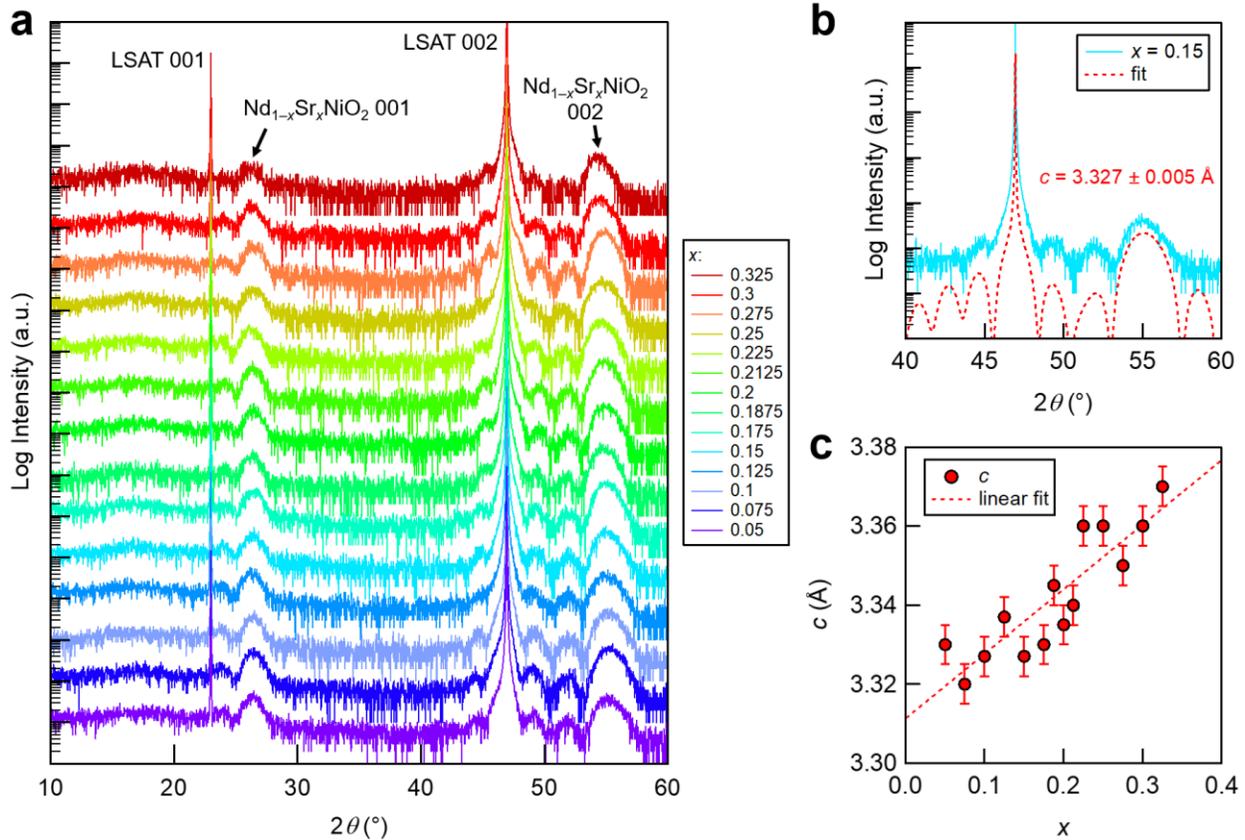

**Extended Data Fig. 1 | X-Ray diffraction of $Nd_{1-x}Sr_xNiO_2$ on LSAT substrates. a,** Representative x-ray diffraction (XRD) $\theta$-$2\theta$ symmetric scans of optimized $Nd_{1-x}Sr_xNiO_2$ ($x$ = 0.05-0.325). The curves are vertically offset for clarity. **b,** XRD $\theta$-$2\theta$ symmetric scan of $Nd_{0.85}Sr_{0.15}NiO_2$ (solid curve) and the corresponding symmetric scan fit (dashed curve). The close agreement in the positions of the main film peak and the Laue fringes indicates a good fit. The extracted out-of-plane lattice constant $c$ from the fit is labeled. **c,** Extracted values of $c$ from XRD $\theta$-$2\theta$ symmetric scan fits plotted against $x$. Error bars are the larger of the error in the fit and standard deviation in the values from multiple samples. $c$ increases linearly with $x$, consistent with systematic doping of Sr in the films.

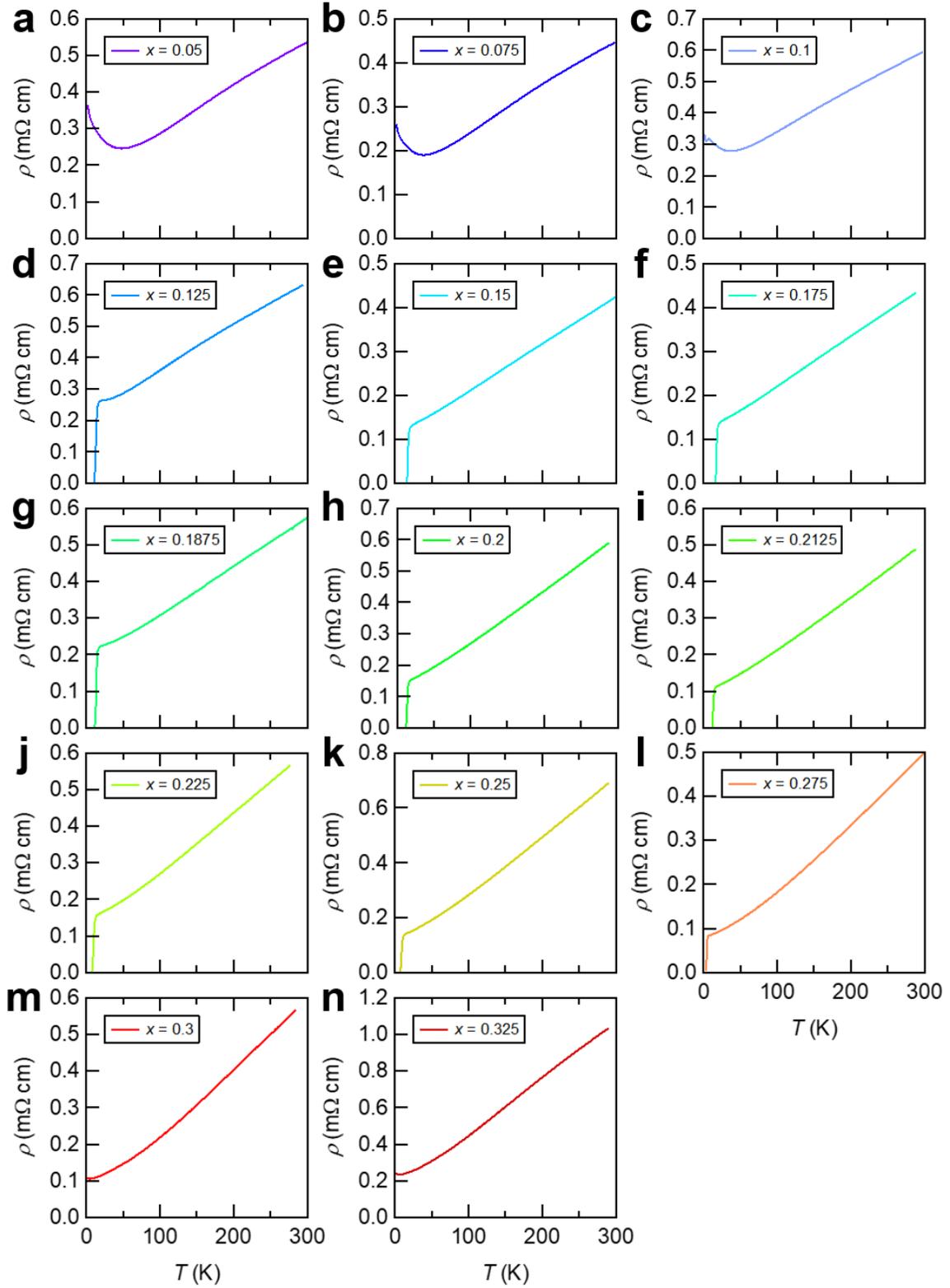

**Extended Data Fig. 2 | Individual resistivity curves of Nd$_{1-x}$Sr$_x$NiO$_2$ on LSAT.** $\rho$ versus $T$ curves of optimized Nd$_{1-x}$Sr$_x$NiO$_2$ films ($x$ = 0.05-0.325).

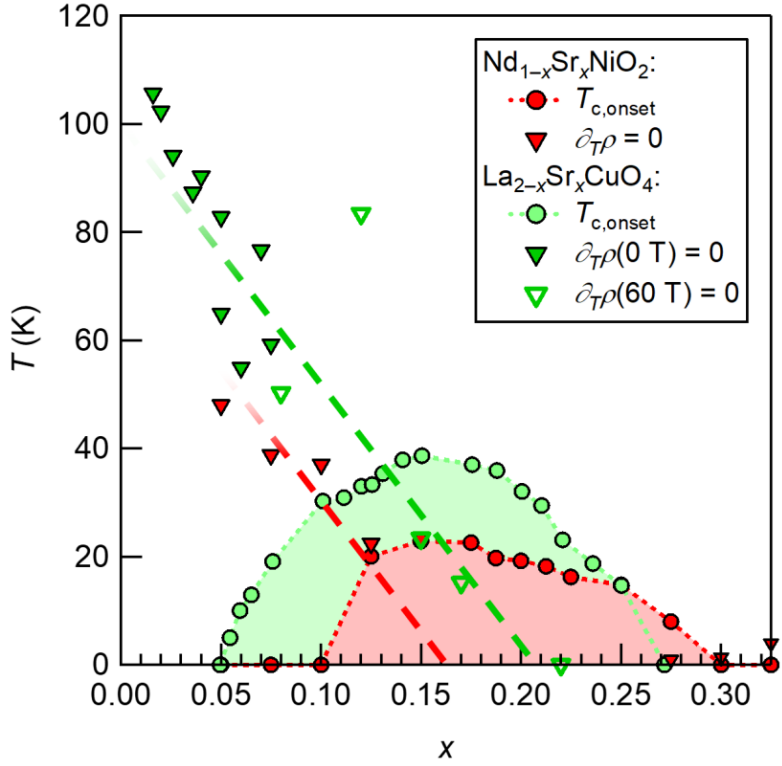

**Extended Data Fig. 3 | Comparing the nickelate and the cuprate phase diagrams.** Superconducting phase diagram of $Nd_{1-x}Sr_xNiO_2$ on LSAT (red) and $La_{2-x}Sr_xCuO_4$ (green).[13,30–32] The superconducting onset temperature is shown via circles, while the resistive upturn temperature $T_{upturn}$ is shown as triangles. For $La_{2-x}Sr_xCuO_4$, the open triangles are $T_{upturn}$ obtained by suppressing superconductivity with high magnetic field.[31] The superconducting dome extends from $x \approx 0.1$-$0.3$ ($\Delta x \approx 0.2$) for $Nd_{1-x}Sr_xNiO_2$ and $x \approx 0.05$-$0.26$ ($\Delta x \approx 0.21$) for $La_{2-x}Sr_xCuO_4$.

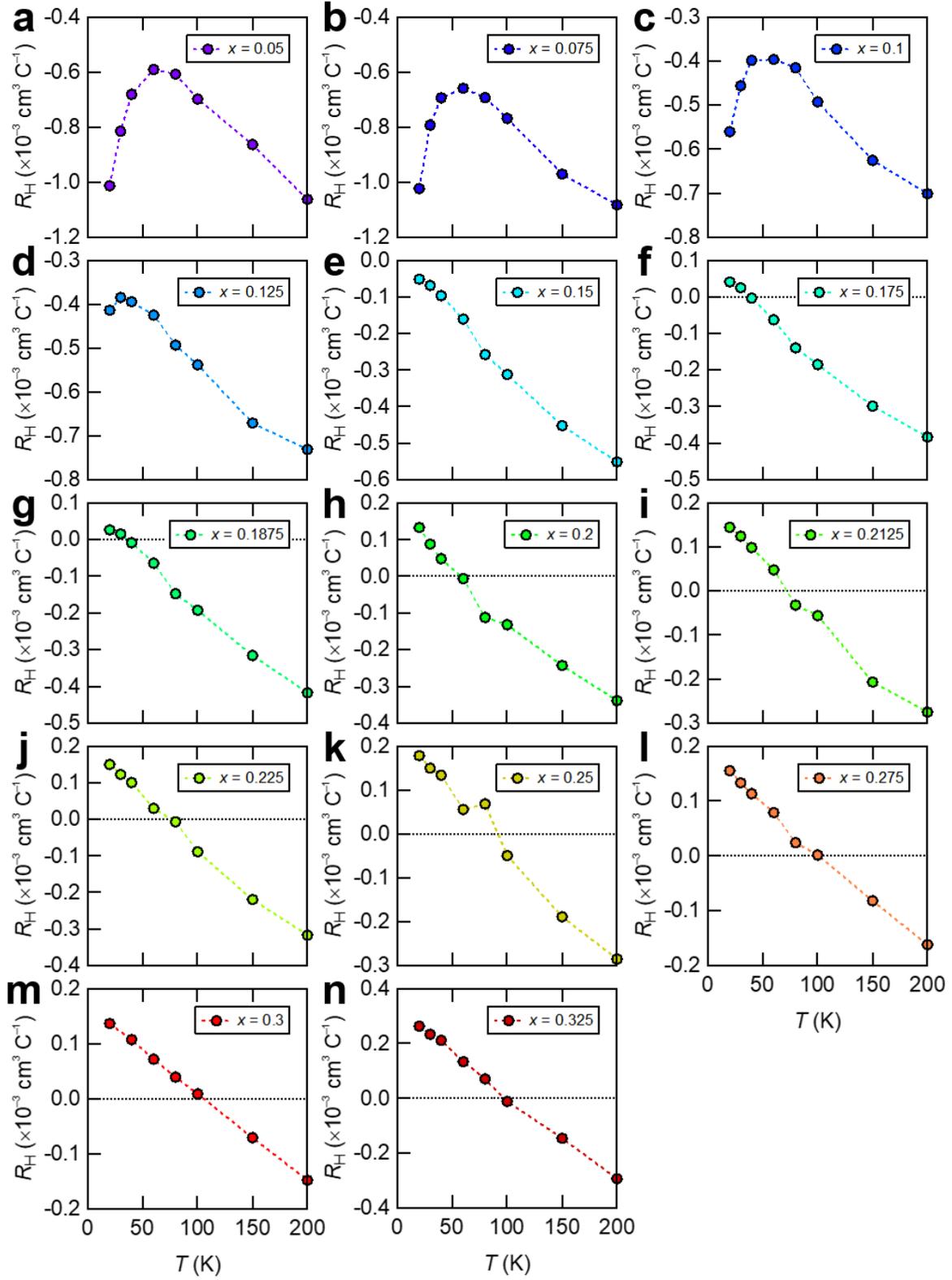

**Extended Data Fig. 4 | Individual $R_H(T)$ curves of Nd$_{1-x}$Sr$_x$NiO$_2$ on LSAT.** $R_H$ versus $T$ curves of optimized Nd$_{1-x}$Sr$_x$NiO$_2$ films ($x$ = 0.05-0.325). $R_H = 0$ is marked as a black dotted line.

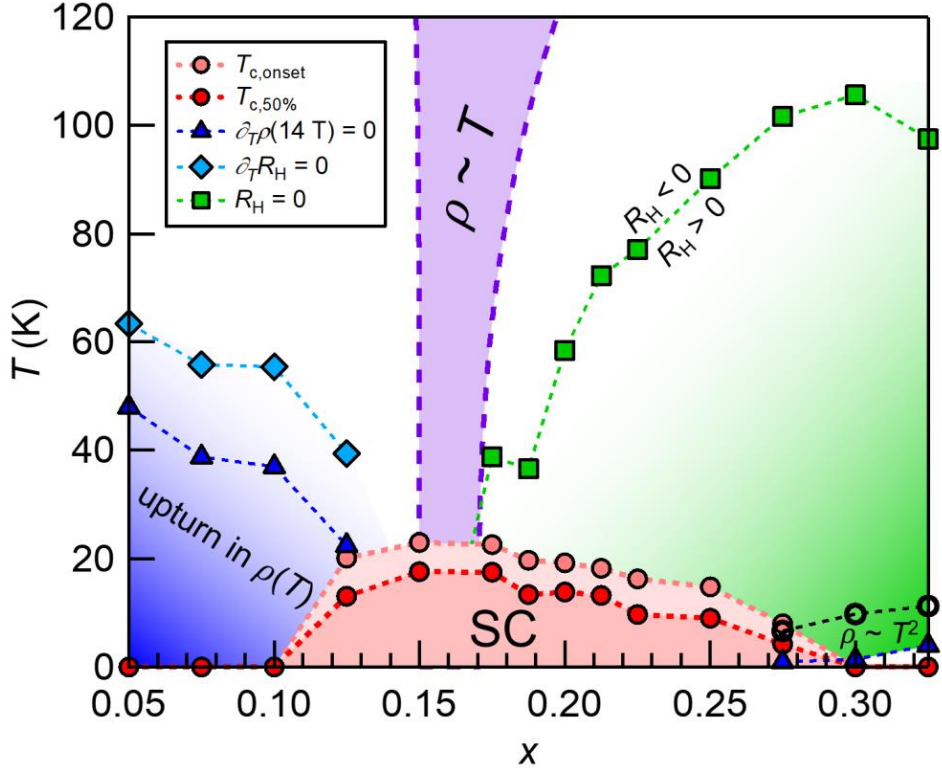

**Extended Data Fig. 5 | Cumulative phase diagram of the infinite-layer nickelate Nd$_{1-x}$Sr$_x$NiO$_2$ on LSAT.** The main features of the phase diagram of Nd$_{1-x}$Sr$_x$NiO$_2$ are summarized here. The onset temperature of the superconducting transition $T_{c,onset}$ is defined as the temperature at which the second derivative of $\rho(T)$ becomes negative, while the 50% transition temperature $T_{c,50\%}$ is defined as the temperature at which $\rho$ is 50% of $\rho(T_{c,onset})$. In the underdoped region, $\rho$ shows a resistive upturn, with $T_{upturn}$ (dark-blue triangles) decreasing as hole doping is increased and superconductivity emerges. Simultaneously, the local maximum in $R_H$ (light-blue diamonds) tracks the doping dependence of the resistive $T_{upturn}$. Superconductivity emerges at $x \approx 0.1$ and persists up to $x \approx 0.3$. In the optimal doping of $x \approx 0.15$-$0.175$, the normal-state resistivity shows a linear $T$-dependence. As superconductivity is suppressed in the overdoped region, $T^2$ resistivity emerges, with a small resistive upturn at low temperatures (dark-blue triangles) driven by disorder. The open circles at the overdoped region delineate the boundary below which the $T^2$ fit shows good agreement with $\rho$. As $x$ is increased, $R_H$ starts to cross zero into positive values (green squares). This transition occurs near the optimal doping, and the zero-crossing temperature increases into the overdoped region.

**Supplementary Information for**

# Character of the "normal state" of the nickelate superconductors

Kyuho Lee[1,2,*], Bai Yang Wang[1,2], Motoki Osada[1,3], Berit H. Goodge[4,5], Tiffany C. Wang[1,6], Yonghun Lee[1,6], Shannon Harvey[1,6], Woo Jin Kim[1,6], Yijun Yu[1,6], Chaitanya Murthy[2], Srinivas Raghu[1,2], Lena F. Kourkoutis[4,5], and Harold Y. Hwang[1,6]

[1] *Stanford Institute for Materials and Energy Sciences, SLAC National Accelerator Laboratory, Menlo Park, CA 94025, USA*
[2] *Department of Physics, Stanford University, Stanford, CA 94305, USA*
[3] *Department of Materials Science and Engineering, Stanford University, Stanford, CA 94305, USA*
[4] *School of Applied and Engineering Physics, Cornell University, Ithaca, NY 14853, USA*
[5] *Kavli Institute at Cornell for Nanoscale Science, Cornell University, Ithaca, NY 14853, USA*
[6] *Department of Applied Physics, Stanford University, Stanford, CA 94305, USA*

___________________
[*]kyuho@stanford.edu

## Interpretation of $R_H$ under a simplified two-band model

For the parent compound of NdNiO$_2$, electronic structure calculations have converged to a multiband description, with large hole-like surfaces of Ni $3d_{x^2-y^2}$ character centered around the Brillouin zone edges and small hybridized electron pockets centered around the $\Gamma$ and A points.[S1–S18] As holes are doped into the system by Sr substitution, the electron pockets are gradually depleted relative to the hole carriers, with theoretical estimates for their complete depletion ranging from 0.05 to > 0.4.[S8–S10] Thus we develop here a two-band interpretation of the temperature dependence of both $R_H$ and $\rho$ ($\rho_{xx}$) which is clearly relevant to the underdoped region.

First, we note that the conventional two-band formula for the longitudinal and Hall resistivity ($\rho_{xx}$ and $\rho_{yx}$, respectively) can be simplified for this system.[S19,S20] Based on the order of magnitude of $R_H$ and $\rho_{xx}$, we estimate that for a magnetic field of 14 T (the highest field used in our measurements), $\omega_c \tau = \mu B \approx 0.002 \ll 1$ (here, $\omega_c$ is the cyclotron frequency, $\tau$ is the relaxation time, $\mu$ is the carrier mobility, and $B$ is the magnetic field). This is also consistent with the experimentally observed linearity of the Hall response up to at least 14 T. Therefore, we discard the higher-order $B$ terms in the magnetoresistivity formula and obtain

$$\rho_{yx} \approx \frac{1}{e} \frac{n_h \mu_h^2 - n_e \mu_e^2}{(n_h \mu_h + n_e \mu_e)^2} B \text{ and} \tag{1}$$

$$\rho_{xx} \approx \frac{1}{e} \frac{1}{n_h \mu_h + n_e \mu_e}, \tag{2}$$

where $n_i$ and $\mu_i$ are the density and mobility for carrier $i$ respectively, with $i$ being either hole ($h$) or electron ($e$). As described in the main text, we observe a low-temperature resistive upturn and a decrease of $R_H$ (i.e. more negative values). Within our conditions and simplified model, this is a result of the temperature dependence of a subset of the four parameters $n_h$, $n_e$, $\mu_h$, and $\mu_e$.

**Table S1.** Temperature derivative of the simplified two-band formula of $R_H$ and $\rho_{xx}$ when only allowing one of the four parameters to vary with temperature.

| $T$-Varying Parameter | $\partial R_H / \partial T$ | $\partial \rho_{xx} / \partial T$ |
|---|---|---|
| $n_h$ | $\dfrac{1}{e}\left[\dfrac{\mu_h^2(n_h\mu_h + n_e\mu_e) + 2\mu_h(n_e\mu_e^2 - n_h\mu_h^2)}{(n_h\mu_h + n_e\mu_e)^3}\right]\dfrac{\partial n_h}{\partial T}$ | $\dfrac{1}{e}\left[\dfrac{-\mu_h}{(n_h\mu_h + n_e\mu_e)^2}\right]\dfrac{\partial n_h}{\partial T}$ |
| $\mu_h$ | $\dfrac{1}{e}\left[\dfrac{2n_h\mu_h(n_h\mu_h + n_e\mu_e) + 2n_h(n_e\mu_e^2 - n_h\mu_h^2)}{(n_h\mu_h + n_e\mu_e)^3}\right]\dfrac{\partial \mu_h}{\partial T}$ | $\dfrac{1}{e}\left[\dfrac{-n_h}{(n_h\mu_h + n_e\mu_e)^2}\right]\dfrac{\partial \mu_h}{\partial T}$ |
| $n_e$ | $\dfrac{1}{e}\left[\dfrac{-\mu_e^2(n_h\mu_h + n_e\mu_e) + 2\mu_e(n_e\mu_e^2 - n_h\mu_h^2)}{(n_h\mu_h + n_e\mu_e)^3}\right]\dfrac{\partial n_e}{\partial T}$ | $\dfrac{1}{e}\left[\dfrac{-\mu_e}{(n_h\mu_h + n_e\mu_e)^2}\right]\dfrac{\partial n_e}{\partial T}$ |
| $\mu_e$ | $\dfrac{1}{e}\left[\dfrac{-2n_e\mu_e(n_h\mu_h + n_e\mu_e) + 2n_e(n_e\mu_e^2 - n_h\mu_h^2)}{(n_h\mu_h + n_e\mu_e)^3}\right]\dfrac{\partial \mu_e}{\partial T}$ | $\dfrac{1}{e}\left[\dfrac{-n_e}{(n_h\mu_h + n_e\mu_e)^2}\right]\dfrac{\partial \mu_e}{\partial T}$ |

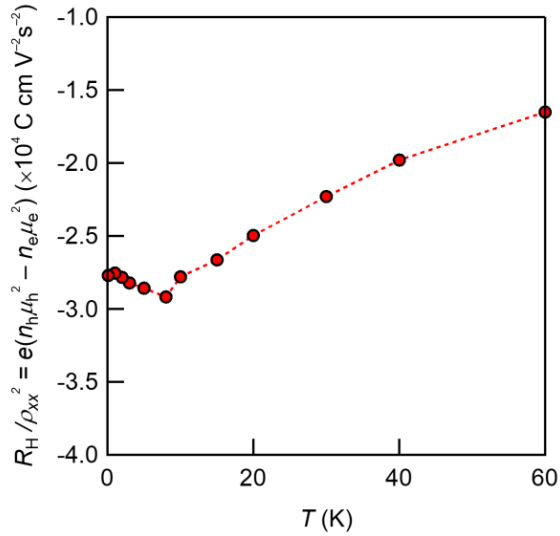

**Fig. S1 | Temperature-dependence of $n_h\mu_h^2 - n_e\mu_e^2$.** Plot of $R_H/\rho_{xx}^2$, which under the simplified two-band model is directly proportional to $n_h\mu_h^2 - n_e\mu_e^2$, for $x = 0.075$. In a considerable range of temperatures over which the resistive upturn occurs, $R_H/\rho_{xx}^2$ increases as function of $T$.

Here, we aim to find a minimal description for such temperature dependence and, therefore, first suppose that only one of the four parameters varies with temperature, with the other three relatively temperature independent. The four possible cases are listed in Table S1. Each row lists the varying parameter and the corresponding temperature derivative of $R_H$ and $\rho_{xx}$. Now, experimentally $R_H$ and $\rho_{xx}$ show opposite trends as function of $T$, namely $\partial R_H/\partial T > 0$ and $\partial \rho_{xx}/\partial T < 0$. However, in the $\mu_e$-varying case, the coefficient in front of $\partial \mu_e/\partial T$ is negative for both $R_H$ and $\rho_{xx}$ (trivially true for $\rho_{xx}$, and for $R_\text{H}$ note that $\frac{n_\text{e}\mu_\text{e}^2 - n_\text{h}\mu_\text{h}^2}{\mu_\text{e}(n_\text{h}\mu_\text{h} + n_\text{e}\mu_\text{e})} < \frac{n_\text{e}\mu_\text{e}^2}{\mu_\text{e}(n_\text{h}\mu_\text{h} + n_\text{e}\mu_\text{e})} < \frac{n_\text{e}\mu_\text{e}^2}{n_\text{e}\mu_\text{e}^2} = 1$), which is inconsistent with the experimental observations. Hence, we eliminate the $\mu_\text{e}$-varying case. We can also rule out the $n_\text{e}$-varying case noting the following. On one hand, the experimental observation of $\partial \rho_{xx}/\partial T < 0$ requires that $\partial n_\text{e}/\partial T > 0$. On the other hand, as shown in Fig. S1 we see that $\partial(n_\text{h}\mu_\text{h}^2 - n_\text{e}\mu_\text{e}^2)/\partial T > 0$ over a considerable region of $T$ at which the resistive upturn occurs, requiring $\partial n_\text{e}/\partial T < 0$. Hence, we reach a contradiction, thus eliminating the $n_\text{e}$-varying scenario.

Importantly, these considerations eliminate both parameters linked to electron carriers and leaves only $n_\text{h}$ and $\mu_\text{h}$ as potential temperature-dependent parameters (within this assumption of having only one temperature dependent contribution). Namely, by this argument, the strong temperature dependence seen in $R_\text{H}$ and $\rho_{xx}$ primarily arises from the hole carrier, and the temperature dependence of the electron carriers likely plays a minor or secondary role.

Having identified the hole carriers to be the origin of the observed temperature dependence, we now aim to distinguish the contributions from its density and mobility variations under the assumption of a single-parameter variation. For this, we utilize the fact that $R_\text{H}$ and $\rho_{xx}$ have identical functional forms below 20 K (Main Text Fig. 3), which is captured by

$$\rho_{xx} \approx -(0.075 \text{ Vs cm}^{-2})R_\text{H} + 0.120 \text{ m}\Omega \text{ cm} \equiv -c_1 R_\text{H} + c_2/2. \tag{3}$$

Taking the temperature derivative of both sides after substituting the simplified two-band formula for $R_\text{H}$ and $\rho_{xx}$, we arrive at

$$-c_1 \partial(\mu_\text{h}^2 n_\text{h})/\partial T + [c_2 e(\mu_\text{h} n_\text{h} + \mu_\text{e} n_\text{e}) - 1]\partial(n_\text{h}\mu_\text{h})/\partial T = 0. \tag{4}$$

**Table S2.** Temperature derivative of the functional-form-matching relation between $R_\text{H}$ and $\rho_{xx}$ and its implications for the two cases of temperature-varying hole density or mobility.

| Assumption | Functional-Form-Matching Relation and Corresponding Implications |
|---|---|
| $\partial n_\text{h}/\partial T \neq 0, \partial \mu_\text{h}/\partial T = 0$ | $[-c_1\mu_\text{h}^2 + c_2 e\mu_\text{h}^2 n_\text{h}(T) + (c_2 e\mu_\text{e} n_\text{e} - 1)\mu_\text{h}]\frac{\partial n_\text{h}(T)}{\partial T} = 0$ <br> $\Rightarrow n_h(T) = \frac{(1-c_2 e\mu_\text{e} n_\text{e})\mu_h + c_1\mu_h^2}{c_2 e\mu_\text{h}^2} \Rightarrow \frac{\partial n_\text{h}}{\partial T} = 0$ (contradiction). |
| $\partial n_\text{h}/\partial T = 0, \partial \mu_\text{h}/\partial T \neq 0$ | $[-c_1\mu_\text{h}(T)n_\text{h} + c_2 e\mu_\text{h}(T)n_\text{h}^2 + (c_2 e\mu_\text{e} n_\text{e} - 1)n_\text{h}]\frac{\partial \mu_\text{h}(T)}{\partial T} = 0$ <br> $\Rightarrow \mu_\text{h}(T)(c_2 e n_\text{h}^2 - c_1 n_\text{h}) = 0; c_2 e\mu_\text{e} n_\text{e} - 1 = 0$ <br> $\Rightarrow c_1 = c_2 e n_\text{h}.$ |

This should hold true for all temperatures at which we observe this functional form matching for $R_H$ and $\rho_{xx}$. Table S2 lists the two possible scenarios, the corresponding equations that need to hold, and their respective implications. For the first assumption of temperature-dependent $n_h$, we arrive at a contradiction. This indicates that $\mu_h$, the only parameter left, would be the one that is temperature dependent. For this to hold, it entails that $c_2 e \mu_e n_e - 1 = 0$, and consequently $c_1 = c_2 e n_h$.

The above analysis yields two implications. First, this suggests that $\mu_h$ is necessarily temperature dependent to describe the low-temperature resistive upturn and the sharp drop of $R_H$. Second, this provides an estimate of the other parameters:

$$\mu_e n_e = \frac{1}{c_2 e} \approx 5.20 \times 10^{22} (\text{V s cm})^{-1} \approx 2.58 \frac{\text{cm}^2}{\text{V s}} \text{ per unit cell volume and} \tag{5}$$

$$n_h = \frac{c_1}{c_2 e} \approx 3.90 \times 10^{21} \text{cm}^{-3} \approx 0.19 \text{ per unit cell volume}, \tag{6}$$

where the unit cell volume is the out-of-plane lattice constant (~3.32 Å, see Extended Data Fig. 2c) multiplied by the square of the LSAT lattice constant (3.868 Å). These numbers are roughly compatible with theoretical expectations of a large hole Fermi surface and small but mobile electron pockets.[S2,S11]

We emphasize that our discussion thus far is based on a simplistic and strong assumption that only one parameter of the four is temperature dependent. However, it is natural to expect that as the density/mobility of one type of carrier varies against temperature, its corresponding mobility/density will also be temperature dependent. Therefore, we now discuss the validity of the above implications if both the density and mobility of only one type of carrier are allowed to be temperature dependent. In this case, electrons alone could explain the observed transport behaviors only if $-2\frac{n_e}{\mu_e}\frac{\partial \mu_e}{\partial T} > \frac{\partial n_e}{\partial T} > -\frac{n_e}{\mu_e}\frac{\partial \mu_e}{\partial T} > 0$, which is indeed not forbidden but a rather very strong, narrow, and unlikely constraint. As it is not obvious that such a constraint holds in general, the hole carrier remains the most likely cause of the transport observations. However, the distinction between the hole mobility and density is significantly weaker under this relaxed assumption. Now, the functional-form-matching condition cannot meaningfully restrict the possible scenarios, and we are left with the broad requirements that $\partial(n_h\mu_h)/\partial T > 0$ and $\partial(n_h\mu_h^2)/\partial T > 0$. In this sense, both the hole mobility and density could contribute to producing the observed resistive upturn and $R_H$ downturn.

In summary, the simplistic two-band model analysis here suggests that the hole carrier is the primary source of the observed low temperature transport behavior. Moreover, a temperature dependent hole mobility is a necessary component in the explanation. The temperature dependence of the hole density cannot be ruled out and instead could be equally important. What appears clear is that the $n_h \mu_h$ product is strongly suppressed with decreasing temperature in the underdoped regime.

## References

S1. Anisimov, V. I., Bukhvalov, D. & Rice, T. M. Electronic Structure of Possible Nickelate Analogs to the Cuprates. *Phys. Rev. B* **59**, 7901 (1999).
S2. Lee, K. W. & Pickett, W. E. Infinite-Layer LaNiO$_2$: Ni$^{1+}$ is not Cu$^{2+}$. *Phys. Rev. B* **70**, 165109 (2004).
S3. Kitatani, M. *et al.* Nickelate Superconductors — a Renaissance of the One-Band Hubbard Model. *npj Quantum Mater.* **5**, 59 (2020).
S4. Been, E. *et al.* Electronic Structure Trends Across the Rare-Earth Series in Superconducting Infinite-Layer Nickelates. *Phys. Rev. X* **11**, 011050 (2021).
S5. Gao, J., Peng, S., Wang, Z., Fang, C. & Weng, H. Electronic Structures and Topological Properties in Nickelates $Ln_{n+1}Ni_nO_{2n+2}$. *Natl. Sci. Rev.* **8**, nwaa218 (2021).
S6. Bernardini, F., Olevano, V. & Cano, A. Magnetic Penetration Depth and $T_c$ in Superconducting Nickelates. *Phys. Rev. Res.* **2**, 013219 (2020).
S7. Ryee, S., Yoon, H., Kim, T. J., Jeong, M. Y. & Han, M. J. Induced Magnetic Two-Dimensionality by Hole Doping in the Superconducting Infinite-Layer Nickelate Nd$_{1-x}$Sr$_x$NiO$_2$. *Phys. Rev. B* **101**, 064513 (2020).
S8. Leonov, I., Skornyakov, S. L. & Savrasov, S. Y. Lifshitz Transition and Frustration of Magnetic Moments in Infinite-Layer NdNiO$_2$ upon Hole Doping. *Phys. Rev. B* **101**, 241108(R) (2020).
S9. Liu, Z. *et al.* Doping Dependence of Electronic Structure of Infinite-Layer NdNiO$_2$. *Phys. Rev. B* **103**, 045103 (2021).
S10. Lechermann, F. Doping-Dependent Character and Possible Magnetic Ordering of NdNiO$_2$. *Phys. Rev. Mater.* **5**, 044803 (2021).
S11. Botana, A. S. & Norman, M. R. Similarities and Differences between LaNiO$_2$ and CaCuO$_2$ and Implications for Superconductivity. *Phys. Rev. X* **10**, 011024 (2020).
S12. Lechermann, F. Late Transition Metal Oxides with Infinite-Layer Structure: Nickelates versus Cuprates. *Phys. Rev. B* **101**, 081110(R) (2020).

S13. Nomura, Y. *et al.* Formation of a Two-Dimensional Single-Component Correlated Electron System and Band Engineering in the Nickelate Superconductor NdNiO$_2$. *Phys. Rev. B* **100**, 205138 (2019).
S14. Werner, P. & Hoshino, S. Nickelate Superconductors: Multiorbital Nature and Spin Freezing. *Phys. Rev. B* **101**, 041104(R) (2020).
S15. Olevano, V., Bernardini, F., Blase, X. & Cano, A. *Ab initio* Many-Body GW Correlations in the Electronic Structure of LaNiO2. *Phys. Rev. B* **101**, 161102(R) (2020).
S16. Karp, J. *et al.* Many-Body Electronic Structure of NdNiO$_2$ and CaCuO$_2$. *Phys. Rev. X* **10**, 021061 (2020).
S17. Wu, X. *et al.* Robust $d_{x^2-y^2}$-Wave Superconductivity of Infinite-Layer Nickelates. *Phys. Rev. B* **101**, 060504(R) (2020).
S18. Sakakibara, H. *et al.* Model Construction and a Possibility of Cupratelike Pairing in a New $d^9$ Nickelate Superconductor (Nd,Sr)NiO$_2$. *Phys. Rev. Lett.* **125**, 077003 (2020).
S19. Allgaier, R. S. High-Field Hall Coefficient in a Compensated, Multiband Conductor. *J. Appl. Phys.* **38**, 5095 (1967).
S20. Ashcroft, N. W. & Mermin, N. D. *Solid State Physics*. (Harcourt, Inc., 1976).